\documentstyle[latexsym,amssymb,aps,floats,eqsecnum,preprint,amsfonts]{revtex}
\def\laq{\raise 0.4ex\hbox{$<$}\kern -0.8em\lower 0.62 ex\hbox{$\sim$}}
\def\gaq{\raise 0.4ex\hbox{$>$}\kern -0.7em\lower 0.62 ex\hbox{$\sim$}}
\def\di{\delta^{({\rm gi})}}
\input epsf
\begin{document}
\draft
\bibliographystyle{unsrt}

\title{Thick branes and Gauss-Bonnet self-interactions}

\author{Massimo Giovannini\footnote{Electronic address: 
Massimo.Giovannini@ipt.unil.ch}}

\address{{\it Institute of Theoretical Physics, 
University of Lausanne}}
\address{{\it BSP-1015 Dorigny, Lausanne, Switzerland}}

\maketitle

\begin{abstract}

Thick branes obtained from a  bulk action containing Gauss-Bonnet 
self-interactions are analyzed in light of the localization properties of 
the various modes of the geometry. 
The entangled system  describing the localization of the tensor, vector 
and scalar fluctuation is decoupled in terms  of variables invariant 
for infinitesimal coordinate transformations. 
The dynamics of the various zero modes is 
discussed and solved in general terms. Provided the four-dimensional 
Planck mass is finite and provided the geometry is everywhere regular,
it is shown that the vector  and scalar zero modes are not localized.
The  tensor zero mode is localized leading to four-dimensional 
gravity. The general formalism is illustrated through specific 
analytical examples.
\end{abstract}
\vskip0.5pc
\centerline{Preprint Number: UNIL-IPT-01-11, July 2001 }
\vskip0.5pc
\noindent
\newpage
\renewcommand{\theequation}{1.\arabic{equation}}
\setcounter{equation}{0}
\section{Introduction} 

If space-time has more than four dimensions, the Einstein-Hilbert term 
is not the only geometrical action leading to equations of motion 
involving (at most) second order derivatives of the metric \cite{lov}.
In  dimensions larger than four, the usual Einstein-Hilbert action can 
indeed be supplemented with higher order curvature corrections 
without generating, in the equations of motion, terms containing 
more than two derivatives of the metric with respect to the 
space-time coordinates \cite{mad}. If this is the case, the quadratic part 
of the action can be written in terms of the Euler-Gauss-Bonnet 
combination:
\begin{equation}
{\cal R}^2_{\rm EGB} = R^{ A B C D} R_{A B C D} - 4 R^{A B} R_{A B} + R^2.
\label{egb}
\end{equation}
In four dimensions the EGB is a topological term and it coincides with 
the Euler invariant: its contribution to the equations of 
motion can be rearranged in a  four-divergence which does not
contribute to the classical equations of motion. In more than 
four dimensions the EGB combination leads to a ghost-free 
theory and it appears in different higher 
dimensional contexts. In string 
theory the EGB indeed appears in the first 
string tension correction to the (tree-level) effective action  
\cite{des,ts,cal,s}.
In supergravity the EGB is required in order to supersymmetrize the 
Lorentz-Chern-Simons term. 

Higher dimensional gravity theories have been investigated in 
connection with possible alternatives to Kaluza-Klein compactification 
\cite{m1,ak,vis,rs,rs2}. For a more complete 
account of the various perspectives of the problem see the recent 
review \cite{rub}.

In this context the gravity part of the action 
is usually taken to be  the Einstein-Hilbert term. 
Recently, various investigations took into account the 
possible contribution of higher derivatives terms in the action and mainly 
in the five-dimensional case \cite{q1,q2,q3,q4}. Various 
physical frameworks can be invoked in order to include 
higher order curvature corrections, whose motivation
 may range from back-reaction 
effects \cite{rg} to some interesting connections 
with string  models \cite{q1}. 
For the r\^ole of quadratic counter-terms in the context of the AdS/CFT 
correspondence see for instance \cite{cft}.
Quadratic corrections may also play a 
role in dimensions larger than five.
In \cite{mg1} seven-dimensional 
warped solutions have been discussed in the case when 
hedgehogs configurations are present together with 
quadratic self-interactions. In \cite{mg2} the simultaneous presence of 
dilaton field and quadratic corrections has been investigated in a 
more string theoretical perspective and mainly in six-dimensions.

Suppose now that a smooth domain-wall solution is used 
in order to localize fields of various spin as in \cite{m1,m2}. 
It is then interesting to investigate how the 
localization of the metric fluctuations (coupled to the 
fluctuations of the wall) is affected by the addition 
of EGB self-interactions. 
Smooth domain wall configurations leading to a 
fully regular geometry can be constructed 
\cite{kt1,kt2,gremm1,gremm2,free,free2}. The warp factors 
interpolates in a regular way between two $AdS_{5}$ 
geometries. If the gravity action is selected to be 
the Einstein-Hilbert term, the scalar and vector zero modes 
of the geometry are not localized on the wall \cite{mg3} while 
the tensor zero mode is localized, leading, ultimately, to ordinary
four-dimensional gravity.
Once quadratic self-interactions  are included in the 
picture, important 
effects on the localization of gravity \cite{kak1} have been discussed 
and it has been also argued that singularity free domain-wall 
solutions are possible 
if the quadratic part of the action is expressed in terms of Gauss-Bonnet 
combination \cite{kak}. 

The aim of the present investigation is to study the fluctuations 
of scalar walls when Gauss-Bonnet self-interactions are included. 
Most of this analysis can be developed in general terms, i.e. 
without specifying the explicit solution describing the domain-wall.
The fluctuations of the metric can be studied in terms 
of fully gauge-invariant
quantities so that the obtained results will be 
independent on the specific coordinate system. Technically this is 
made possible by generalizing the {\em Bardeen formalism} \cite{bar} 
to the case of non-compact extra dimensions whose dynamics is 
described in terms of a specific quadratic theory of gravity of the type of
 the one recalled in Eq. (\ref{egb}). The application of the present 
formalism to the context of large (but compact) extra-dimensions 
\cite{ced1,ced2,ced3} will not be directly examined here.

The plan of the present investigation is then the following. 
In Section II the 
basic equations describing the background dynamics will be presented. 
In Section III the fully gauge-invariant approach to linearized 
fluctuations will be discussed and applied to the problem 
at hand. Section IV deals
the localization of the tensor fluctuations
 whereas in Section V the localization of the vector and scalar 
fluctuations will be analyzed. Section VI contains some specific studies of 
the localization of metric fluctuations in the case of analytical and
singularity-free thick 
brane solutions induced by Gauss-Bonnet self-interactions. 
Some concluding remarks are presented in Section VII. 
Various technical results are collected in the Appendix.

\renewcommand{\theequation}{2.\arabic{equation}}
\setcounter{equation}{0}
\section{Brane sources with quadratic gravity in the bulk}

Consider a bulk action 
containing EGB self-interactions together with a brane 
source describing the spontaneous 
breaking of $D$-dimensional Poincar\'e invariance:
\begin{equation}
S = \int d^D x \sqrt{|G|} \biggl[ - \frac{R}{ 2 \kappa} - \alpha' 
{\cal R}^2_{\rm EGB} + \frac{1}{2} G^{A B} \partial_{A} \varphi 
\partial_{B} \varphi - V( \varphi) \biggr],
\label{ac}
\end{equation} 
where $ \kappa = 8 \pi G_{D}$ is  related to the 
$D$-dimensional Planck mass.
The sign of the coupling appearing in front of the EGB combination 
has been taken in order to match the sign obtained from the 
low-energy string effective action. 
Dimensionally, $[\alpha'] = L^{D -4}$ where $L$ is a generic 
length scale. The potential $V(\varphi)$ will be assumed 
to be symmetric for $\varphi \rightarrow - \varphi$. For sake 
of simplicity, one can think of  the case where 
the potential is
\begin{equation} 
V(\varphi) = V_0 \biggl[ c_1 \biggl( \frac{\varphi}{\varphi_0} \biggr)^4 
+ c_2 \biggl( \frac{\varphi}{\varphi_0} \biggr)^2  + c_3\biggr],
\label{pot}
\end{equation}
where $c_1$, $c_2$ and $c_3$ are numerical coefficients of order one 
(no hierarchy is assumed among them). 
The equations of motion derived from the action of Eq. (\ref{ac})  
\begin{eqnarray}
&& R_{A}^{B} - \frac{1}{2} \delta_{A}^{B} R = \kappa T_{A}^{B} - 
2 \alpha' \kappa
{\cal Q}_{A}^{B},
\label{e1}\\
&& G^{ A B} \nabla_{A} \nabla_{B} \varphi + 
\frac{\partial V}{\partial\varphi} =0,
\label{s1}
\end{eqnarray}
are written in terms of the brane energy-momentum tensor 
\begin{equation}
T_{A}^{B} = \partial_{A} \varphi \partial^{B} \varphi - \delta_{A}^{B} 
\biggl[ \frac{1}{2}G^{M N} 
\partial_{M} \varphi \partial_{N} \varphi - V(\varphi) \biggr],
\label{e1a}
\label{tens} 
\end{equation}
and of the Lanczos tensor
\begin{equation}
{\cal Q}_{A}^{B} = \frac{1}{2} \,\delta_{A}^{B}\, {\cal R}^2_{\rm EGB} -
2\,R\, R_{A}^{B} + 4\,R_{A\,C} \, R^{C\,B} + 4\, R_{CD}\,R_{A}^{~~C\,B\,D}
-\,R_{A\,C\,D\,E}\, R^{B\,C\,D\,E}, 
\label{e1b}
\end{equation}
 accounting for the contribution of quadratic corrections 
to the equations of motion. 

In the contracted form of Eqs. (\ref{e1})--(\ref{e1b})
\begin{equation}
R_{A}^{B} = \kappa \tau_{A}^{B}\, - \,\epsilon \,\biggl[ \delta_{A}^{B} 
\frac{1}{d} \, {\cal R}^2_{\rm EGB}\, - 2 \, R\, R_{A}^{B} + 
4\, R_{A\,C} \, R^{C\,B} + 4 \, R_{C\, D}R_{ A}^{~~C\,B\, D} - 
2\, R_{A\,C\,D\,E} \, R^{B\,C\,D\,E} \biggr],
\label{e2}
\end{equation}
we set, for notational convenience, $\epsilon = 2 \alpha' \kappa$ and 
$ d = D -2$. In Eq. (\ref{e2}) 
\begin{equation}
\tau_{A}^{B} = T_{A}^{B} - \frac{T}{d} \delta_{A}^{B},
\end{equation}
and $T\equiv T_{A}^{A}$ is  the trace of the brane 
energy-momentum tensor. 
The $D$-dimensional 
metric will be taken in the form 
\begin{equation}
ds^2 = a^2(w) [ dt^2 - d\,x_1^2 -\,...-d\,x_{d}^2 - d\,w^2],
\label{metric}
\end{equation}
where the ellipses stand for the $d$ spatial coordinates on the brane 
and while $w$ is the bulk coordinate \footnote{The Latin (uppercase) indices 
run over the whole $D$-dimensional space-time whereas the Greek indices 
run over the $(d+ 1)$-dimensional  subspace.}. 

Defining 
${\cal H} = a'/a$, in the metric of Eq. (\ref{metric}) the EGB 
combination can be written as
\begin{equation}
{\cal R}^2_{\rm EGB} = \frac{d ( d + 1)}{a^4} \biggl[ ( d -1 )\, (d -2)\, 
{\cal H}^4 + 4 ( d -1) {\cal H}'\,{\cal H}^2 \biggr],
\label{gbex}
\end{equation}
where the prime denotes the derivation with respect to the 
bulk coordinate $w$.
The non-vanishing components of the Lanczos tensor are 
\begin{eqnarray}
&& {\cal Q}_{\mu\,\nu} = \frac{1}{2 a^2} \biggl[ d\,(d -1)\,(d-2)\,(d-3) \,
{\cal H}^4 \, + 4 \,d\, ( d-1) \, (d-2) {\cal H}^2 {\cal H}' \biggr] 
\eta_{\mu\nu},
\label{q1}\\
&& {\cal Q}_{w\,w} = - \frac{1}{2 a^2} d\,(d+1)\,(d-1) \, (d-2) \,{\cal H}^4, 
\label{q2}
\end{eqnarray}
where $\eta_{\mu\,\nu}$ is the Minkowski metric in $(d+1)$-dimensions. 

Using Eqs. (\ref{gbex})--(\ref{q2}) the explicit form of Eqs. 
(\ref{e1})--(\ref{s1}) can be obtained for a metric of the type 
of the one reported in Eq. (\ref{metric}):
\begin{eqnarray}
&& d\, {\cal H}' + \frac{d( d-1)}{2} {\cal H}^2 = 
- \kappa \biggl[ \frac{{\varphi'}^2}{2} + V a^2 \biggr] 
\nonumber\\
&+&
 \frac{\epsilon}{a^2} \biggl[ \frac{d\,(d-1)\,(d-2)\,(d-3)}{2} {\cal H}^4 
+ 2\,d\,(d-1)\,(d-2) {\cal H}^2 {\cal H}' \biggr],
\label{ex1}\\
&& \frac{d\,(d+ 1)}{2} {\cal H}^2 = \kappa \biggl[ \frac{{\varphi'}^2}{2} 
- V a^2 \biggr] + \frac{\epsilon}{2 \,a^2} d\,(d +1)\,(d-1)\,(d-2) {\cal H}^4,
\label{ex2}\\
&& \varphi'' + d\, {\cal H} \varphi' - \frac{\partial V}{\partial \varphi} 
a^2 =0.
\label{ex3}
\end{eqnarray}
By combining Eqs. (\ref{ex1})--(\ref{ex3}) we obtain
\begin{eqnarray}
&& {\varphi'}^2 = \frac{d}{\kappa} ( {\cal H}^2 - {\cal H}') q(w),
\label{vph}\\
&& V + \frac{ d} { 2 \kappa a^2} \biggl\{ ({\cal H}' + d {\cal H}^2) 
- \frac{\epsilon}{a^2} (d-1) (d -2) [ ( d-1) {\cal H}^4 + 2 {\cal H}^2 
{\cal H}']\biggr\}=0,
\label{V}
\end{eqnarray}
where 
\begin{equation}
q(w)= \biggl[1 - \frac{2 \epsilon}{a^2} (d-1)(d-2) {\cal  H}^2 \biggr]
\label{q}
\end{equation}
has been defined since it  naturally appears in the analysis 
of the metric fluctuations.

The relation among the reduced 
Planck mass and the higher dimensional Planck mass is modified with 
respect to the case when quadratic curvature corrections are absent. 
In the interesting  case when $d=3$  
\begin{equation}
M_{\rm P}^2 \simeq M^3 \int_{0}^{\infty} a^3(w)\biggl[ 1 
+ \frac{4\epsilon}{a^2} 
({\cal H}^2 + 2 {\cal H}') \biggr] dw.
\label{rel1}
\end{equation}
In the case when $\epsilon =0$, Eq. (\ref{rel1}) turns into 
the the well know relation connecting 
the four-dimensional Planck mass to the five-dimensional one.

\renewcommand{\theequation}{3.\arabic{equation}}
\setcounter{equation}{0}
\section{ Gauge-invariant theory of linearized fluctuations}
When EGB self-interactions are included in the 
action the theory of linearized fluctuations 
is more complicated than in the case where the bulk action only 
contains the Einstein-Hilbert term. The system of equations 
describing the fluctuations can be written, formally, as 
\begin{equation}
\delta R_{A}^{B} = \kappa \delta\tau_{A}^{B} - 2\alpha' 
\kappa\biggl[  \delta {\cal Q}_{A}^{B} - 
\frac{1}{d} \delta {\cal Q}~ \delta_{A}^{B}\biggr],
\label{fl1}
\end{equation}
with
\begin{equation}
\delta \tau_{A}^{B} = \partial_{A}\varphi\partial^{B} \chi + 
\partial_{A}\chi \partial^{B} \varphi -
\frac{2}{d} \delta_{A}^{B} \frac{\partial V}{\partial\varphi} \chi,
\label{fl2}
\end{equation}
and where  $\chi$ denotes the fluctuation of $\varphi$. The 
fluctuations of the Lanczos tensor (and of its trace) appearing 
in Eq. (\ref{fl1}) can be written as  
\begin{eqnarray} 
 \delta {\cal Q}_{A}^{B} &=& \biggl[\frac{1}{2} \delta {\cal R}^2_{\rm EGB} 
\delta_{A}^{B} - 2 \overline{R}~ 
\delta R_{A}^{B}  - 2 \delta R~ \overline{R}_{A}^{B} +  4 \delta R_{A C}~
\overline{R}^{C B}
+ 4\, \overline{R}_{ A C}~\delta R^{CB}
\nonumber\\
 &+& 4 \delta R_{C D} ~\overline{R}_{A}^{~~CBD} 
+ 4 \overline{R}_{C D} ~\delta R_{A}^{~~CBD}  - 2 \delta R_{A C D E} 
\overline{R}^{B C D E}
- 2 \overline{R}_{A C D E}~ \delta R^{B C D E}\biggr],
\label{fl3}\\
 \delta {\cal Q} &=&\biggl(\frac{d - 2}{2} \biggr) 
\delta {\cal R}^2_{\rm EGB} 
\equiv  ( d - 2) \biggl[ \overline{R} 
\delta R - 2 \overline{R}_{M N} \delta R^{MN}
- 2 \delta R_{MN} \overline{R}^{MN} 
\label{fl4}\\
 &+& \frac{1}{2} \overline{R}_{MNAB} \delta 
R^{MNAB}  + \frac{1}{2} 
\delta R_{MNAB} \overline{R}^{MNAB} \biggr],
\\
 \delta R &=& \overline{R}_{M N} \delta G^{M N} + \overline{G}_{M N} \delta 
R^{M N},
\label{fl5}
\end{eqnarray}
where the symbol $\delta$ indicates the fluctuations of the various tensors 
to first order in the amplitude of the metric fluctuations
\begin{equation}
G_{A}^{B} ( x^{\mu}, w) = 
\overline{G}_{A}^{B}(w) + \delta G_{A}^{B}(x^{\mu},w),
\end{equation}
and where the over-line reminds that the corresponding 
quantities are evaluated on the background.
Eqs. (\ref{fl1})--(\ref{fl5}) should be supplemented with 
\begin{equation}
\delta G^{A B} ( \partial_{A} \partial_{B} \varphi - 
\overline{\Gamma}_{A B }^{C} \partial_{C} \varphi) + \overline{G}^{A B} 
(\partial_{A } \partial_{B} \chi - \overline{\Gamma}_{A B }^{C} \partial_{C}
 \chi - 
\delta \Gamma_{A B }^{C} \partial_{C} \varphi) 
+ \frac{\partial V}{\partial\varphi} \chi =0,
\label{fl6}
\end{equation}
which is the perturbed counterpart of Eq. (\ref{s1}).

The gauge-invariant theory of linearized fluctuations 
\cite{mg3,bar} can be generalized to the case when the 
gravity action is not in the Einstein-Hilbert form. 
Even if the metric fluctuations are not, by themselves, 
invariant under infinitesimal 
coordinate transformations,  fully 
gauge-invariant equations can be obtained using a two-step 
procedure which will be now illustrated.

Eqs. (\ref{fl1})--(\ref{fl6}) should be written in general 
terms without choosing a specific coordinate system but keeping 
all the $(d+2)(d+3)/3$ ($15$ for $d=3$) degrees of freedom of the metric. 
Then the variation can be written in fully gauge-invariant terms by selecting 
the appropriate variables invariant under infinitesimal coordinate 
transformations.
This procedure is strongly reminiscent of what is done in the context of the 
Bardeen formalism \cite{bar} and of its generalization to the case 
of compact Kaluza-Klein dimensions \cite{ab,mg4}.

The second step will be to write down, explicitly, Eqs.
(\ref{fl1})--(\ref{fl6}) in terms of the gauge-invariant fluctuations and 
to decouple the system. In order to complete this second step the 
background equations (\ref{ex1})--(\ref{ex3}) will be used
into the perturbed equations together with the constraints 
arising from the off-diagonal components of the perturbed equations.

\subsection{Fully gauge invariant approach}

Without assuming any specific gauge the fluctuations of the 
Ricci tensor can be obtained, after a tedious calculation:
\begin{eqnarray}
\delta R_{w}^{w} &=& \frac{(d + 1)}{a^2}\biggl\{ \psi'' + {\cal H} (\xi' 
+ \psi' ) + 2 {\cal H}' \xi 
\nonumber\\
&-& \partial_{\alpha} \partial^{\alpha}[ ( C - E')' 
+ {\cal H} ( C - E') - \xi]\biggr\}, 
\label{p1}\\
\delta R_{\mu}^{w} &=& \frac{1}{a^2} \biggl\{ d \partial_{\mu} ( {\cal H} \xi 
+ \psi') 
+ \frac{1}{2} \partial_{\alpha}\partial^{\alpha} ( D_{\mu} - 
f_{\mu}') \biggr\},
\label{p2}\\
\delta R_{\mu}^{\nu} &=& \frac{1}{a^2} \biggl\{ {h_{\mu}^{\nu}}'' + 
d {\cal H} {h_{\mu}^{\nu}}'  - \partial_{\alpha}\partial^{\alpha} 
h_{\mu}^{\nu} 
\nonumber\\
&+&\delta_{\mu}^{\nu} \bigl[ \psi'' + ( 2 d + 1) {\cal H} \psi' 
- \partial_{\alpha} \partial^{\alpha} \psi 
+ {\cal H} \xi' + 2 ({\cal H}' + d {\cal H}^2 ) \xi - 
{\cal H} \partial_{\alpha}\partial^{\alpha} ( C - E') \bigr]  
\nonumber\\
&+&  
\partial_{\mu} \partial^{\nu}[ (E' - C)' + d {\cal H} (E' - C) + \xi
- (d -1) \psi]
\nonumber\\
&+& [ (\partial_{(\mu}f^{\nu)})'' + d {\cal H}  (\partial_{(\mu}f^{\nu)})']
- [ (\partial_{(\mu}D^{\nu)})' + d {\cal H}  (\partial_{(\mu}D^{\nu)})]
\biggr\},
\label{p3}
\end{eqnarray}
where the various functions appearing in the fluctuations come from the 
perturbed form of the metric which has been taken as 
\begin{equation}
\delta G_{A B}=a^2(w) \left(\matrix{2 h_{\mu\nu} 
+(\partial_{\mu} f_{\nu} +\partial_{\nu} f_{\mu}) 
+ 2\eta_{\mu \nu} \psi
+ 2 \partial_{\mu}\partial_{\nu} E
& D_{\mu} + \partial_{\mu} C &\cr
D_{\mu} + \partial_{\mu} C  & 2 \xi &\cr}\right).
\label{pm}
\end{equation}
In Eq. (\ref{pm}) $h_{\mu\nu}$ is a divergence-less and trace-less 
rank-two tensor in the $(d+1)$-dimensional Poincar\'e invariant 
space-time. The vectors $f_{\mu}$ and $D_{\mu}$ are both divergence-less.
The scalar fluctuations are parametrized by the four independent 
functions $ \xi$, $\psi$, $C$ and $E$. 
The number of independent functions parameterizing the fluctuations 
of the metric is then $(d+2)(d+3)/2$.

Under an infinitesimal coordinate transformation of the form
\begin{equation}
x^{A} \rightarrow \tilde{x}^{A} = x^{A} + \epsilon^{A},
\label{trans}
\end{equation}
the fluctuations of the metric transform according to the 
usual expression involving the Lie derivative in the direction 
of the vector $\epsilon^{A}$
\begin{equation}
\delta \tilde{G}_{A B} = \delta G_{AB} - \nabla_{A} \epsilon_{B} - \nabla_{B}
\epsilon_{A},
\label{liederiv}
\end{equation}
where $\epsilon_A = a^2(w) (\epsilon_{\mu}, -\epsilon_{w})$ and where 
the gauge functions can be written as 
\begin{equation}
\epsilon_{\mu} = \partial_{\mu} \epsilon + \zeta_{\mu}
\end{equation}
with $\partial_{\mu} \zeta^{\mu} =0$.

As defined in Eq. (\ref{pm}), the tensor modes of the metric fluctuations, 
i.e. $h_{\mu\nu}$ are automatically invariant under the transformation 
(\ref{trans}), whereas  
the vectors and the scalars are not gauge-invariant. This 
is the source of the lack of gauge-invariance of Eqs. (\ref{p1})--(\ref{p3}).
However, since there are two scalar gauge functions  [i.e. 
$\epsilon$ and $\epsilon_{w}$],
and one vector gauge function [i.e. $\zeta_{\mu}$], two
gauge-invariant scalar fluctuations  and one gauge-invariant vector fluctuation
can be constructed. The gauge-invariant scalars are
\begin{eqnarray}
&&\Psi = \psi - {\cal H}  ( E' - C), 
\label{giscal0}\\
&& \Xi = \xi - \frac{1}{a} [ a( C - E')]'.
\label{giscal}
\end{eqnarray}
The gauge-invariant vector is
\begin{equation}
V_{\mu} = D_{\mu} - f_{\mu}'.
\label{givec}
\end{equation}
The invariance of Eqs. (\ref{giscal0})--(\ref{givec}) with respect to 
infinitesimal 
coordinate transformations can be verified by using the explicit form 
of Eq. (\ref{liederiv}), namely:
\begin{equation}
\tilde{h}_{\mu\nu} = h_{\mu\nu},
\label{hl}
\end{equation}
for the tensors and 
\begin{eqnarray}
&& \tilde{f}_{\mu} = f_{\mu} - \zeta_{\mu},
\label{fl}\\
&&\tilde{D}_{\mu} = D_{\mu} - \zeta_{\mu}'.
\label{zeta}
\end{eqnarray}
for the vectors.
Since according to Eq. (\ref{liederiv}) 
the scalars transform as
\begin{eqnarray}
&&\tilde{E} = E - \epsilon,
\label{El}\\
&&\tilde{\psi} = \psi - {\cal H} \epsilon_{w},
\label{psil}\\
&& \tilde{C} = C - \epsilon' + \epsilon_{w},
\label{Cl}\\
&& \tilde{\xi} = \xi + {\cal H} \epsilon_{w} + \epsilon_{w}'.
\label{xil}
\end{eqnarray}
the invariance of Eqs. (\ref{giscal0})--(\ref{giscal}) 
is immediately verified. 
Sometimes, in cosmological applications, the gauge-invariant 
fluctuations are called Bardeen potentials.
 
Using Eqs. (\ref{giscal0})--(\ref{givec}) into Eqs. (\ref{p1})--(\ref{p3}) 
the fluctuations of the Ricci tensor can be written as the sum of 
a fully gauge-invariant part and of a second term which will 
vanish using the equations of the background:
\begin{eqnarray}
&&\delta R_{w}^{w} = \delta^{{\rm (gi)}} R_{w}^{w} - \bigl[\overline{R}_{w}^{w}
\bigr]'( C - E'), 
\label{g1}\\
&& \delta R_{\mu}^{w} 
= \delta^{{\rm (gi)}} R_{\mu}^{w} - \overline{R}_{w}^{w} 
\partial_{\mu}( C - E') + \overline{R}_{\mu}^{\nu} \partial_{\nu} ( C - E'),
\label{g2}\\
&&  \delta R_{\mu}^{\nu} 
= \delta^{{\rm (gi)}} R_{\mu}^{\nu} - [\overline{R}_{\mu}^{\nu}]' 
( C - E'),
\label{g3}
\end{eqnarray}
where $\delta^{{\rm (gi)}}$ denotes a variation which 
preserves gauge-invariance and where
\begin{eqnarray}
\delta^{{\rm (gi)}} R_{w}^{w}   &=& \frac{1}{a^2} \biggl\{ (d + 1)
[\Psi'' + {\cal H} ( \Psi' + \Xi') + 2 {\cal H}' \Xi] + 
\partial_{\alpha}\partial^{\alpha} \Xi\biggr\},
\label{gir1}\\
\delta^{{\rm (gi)}} R_{\mu}^{w} &=& \frac{1}{a^2}\biggl\{ d \partial_{\mu}[
{\cal H} \Xi + \Psi'] + \frac{1}{2} \partial_{\alpha}\partial^{\alpha} V_{\mu}
\biggr\},
\label{gir2}\\
 \delta^{{\rm (gi)}} R_{\mu}^{\nu} &=& \frac{1}{a^2} \biggl\{  
\delta_{\mu}^{\nu} [ \Psi'' + ( 2 d + 1) {\cal H} \Psi' - 
\partial_{\alpha} \partial^{\alpha} \Psi + {\cal H} \Xi' + 
2 ( {\cal H}' + d {\cal H}^2) \Xi] 
\nonumber\\
&+& \partial_{\mu} \partial^{\nu} [ \Xi 
- ( d - 1) \Psi]  - [ (\partial_{(\mu}V^{\nu)})' + d {\cal H} 
(\partial_{(\mu} V^{\nu)})] +{h_{\mu}^{\nu}}''
+ d {\cal H} {h_{\mu}^{\nu}}' - \partial_{\alpha} \partial^{\alpha} 
h_{\mu}^{\nu}  \biggr\}.
\label{gir3}
\end{eqnarray}

The same procedure outlined in the case of the Ricci tensors should
be repeated for all the tensors appearing in Eqs. (\ref{fl1})--(\ref{fl6}).
For sake of simplicity, in view of the rather heavy algebraic 
expressions, the tensor, vector and scalar modes will 
be separately discussed. Moreover, the relevant technical results will 
be collected in the various sections of the Appendix.

Under infinitesimal coordinate transformations also the scalar 
field fluctuation $\chi$ changes as 
\begin{equation}
\tilde{\chi}= \chi - \varphi' \epsilon_{w},
\end{equation}
and the corresponding gauge-invariant variable is 
\begin{equation}
X = \chi - \varphi' (E' - C).
\end{equation}
Also for the fluctuations of the brane source the same procedure 
outlined for the Ricci tensors should be performed  so that fully 
gauge-invariant fluctuations can be obtained.

\renewcommand{\theequation}{4.\arabic{equation}}
\setcounter{equation}{0}
\section{Localization of tensor modes}

Taking the contracted 
form of Einstein equations into account, Eqs. 
(\ref{fl1})--(\ref{fl6}) perturbed to first order in the amplitude 
of (gauge-invariant) tensor fluctuations $h_{\mu}^{\nu}$ can be written as
\begin{eqnarray}
&& \biggl[ 1 - 2 \epsilon \overline{R} \biggr]~ \delta R_{\mu}^{\nu} + 
2 \epsilon\biggl\{2 [\delta R_{ \alpha \beta }~ 
\overline{R}_{\mu}^{~~\alpha\nu \beta} + \overline{R}_{\alpha\beta}~
\delta R_{\mu}^{~~\alpha \nu \beta}+ \overline{R}_{ww}~\delta 
R_{\mu}^{~~w \nu w} ]
\nonumber\\
&&-  [ \delta R_{\mu C D E}~\overline{R}^{\nu C D E} + 
\overline{R}_{\mu C D E}~\delta R^{ \nu C D E} ] +
2 [  \delta R_{\mu \alpha}~\overline{R}^{\alpha \nu} +
\overline{R}_{\mu \alpha}~\delta R^{ \alpha \nu} ]  \biggr\}=0.
\label{peq}
\end{eqnarray}
where the symmetries of the background (i.e. $\overline{R}_{\mu w}=0$, etc.) 
have been used, when possible, in order to simplify the contractions.
Recalling now the explicit form of the background tensors reported in Eqs. 
(\ref{r1})--(\ref{r4}) and their perturbed version reported in 
Eqs. (\ref{ptr1})--(\ref{ptr3})  
of Appendix B the following equation can be obtained 
\begin{eqnarray}
&&\biggl\{ 1 - \frac{2 \epsilon}{ a^2} {\cal H}^2 ( d-1)\,(d-2) \biggr\} 
{h_{\mu}^{\nu}}'' 
\nonumber\\
&&+ \biggl\{ d {\cal H} - \frac{ 2 \epsilon {\cal H}}{a^2} 
(d-1)\,(d-2) [ 2 {\cal H}' + (d-2) {\cal H}^2] {h_{\mu}^{\nu}}'\biggr\}
{h_{\mu}^{\nu}}'
\nonumber\\
&&  -
\biggl\{ 1 - \frac{2 \epsilon}{a^2} (d -2) \biggl[ 2 {\cal H}' + ( d- 3) 
{\cal H}^2 \biggr] \biggr\} \partial^{\alpha}\partial_{\alpha} h_{\mu}^{\nu} 
=0.
\label{graveq}
\end{eqnarray}
The different contributions appearing in Eq. (\ref{peq}) and leading to 
Eq. (\ref{graveq}) are 
listed in Eqs. (\ref{peq1})--(\ref{peq3}) of Appendix B.
Using now the explicit definition of $q(w)$ reported in Eq. (\ref{q}), 
Eq. (\ref{graveq}) becomes
\begin{equation}
q {h_{\mu}^{\nu}}'' + ( d {\cal H} q + q' ) {h_{\mu}^{\nu}}' - 
\biggl[ q + \frac{ q'}{( d - 1 ) {\cal H}}\biggr] 
\partial_{\alpha} \partial^{\alpha} h_{\mu}^{\nu} =0, 
\label{graveq2}
\end{equation}
where the relation 
\begin{equation}
q' = \frac{ 4 \epsilon}{a^2} ( d-1) ( d- 2) {\cal H} ( {\cal H}^2 - {\cal H}').
\end{equation}

By eliminating first derivative, Eq. (\ref{graveq2}) can be finally 
put in a Schr\"odinger-like form, namely 
\begin{equation}
\mu'' - \frac{({\sqrt{s}})''}{\sqrt{s}} \mu - \frac{r}{s} \partial_{\alpha} 
\partial^{\alpha}\mu=0.
\label{canvar}
\end{equation}
where, for a generic tensor polarization,
\begin{equation}
\mu (w, x^{\mu}) = \sqrt{s(w)} h(x^{\mu}, w), 
\end{equation}
and where 
\begin{eqnarray}
&& s(w) = a^{d} \biggl[1 - 2 \epsilon (d-1) (d-2) 
\frac{{\cal H}^2}{a^2}\biggr]\equiv a^{d} q,
\nonumber\\
&& r(w) = a^{d} \biggl\{ 1 - 2 \frac{\epsilon (d - 2) }{ a^2}  \biggl[ 
2 {\cal H}'  + ( d - 3) {\cal H}^2\biggr] \biggr\} \equiv a^{d}
\biggl( q + \frac{ q'}{( d - 1 ) {\cal H}}\biggr).
\end{eqnarray}

Not surprisingly the same equation can be directly obtained by perturbing the 
action (\ref{ac}) to second order in the amplitude of the tensor modes with
the result that, up to total derivatives,  
\begin{equation}
\delta^{(2)} S= \frac{1}{2} \int d^{d + 2} x \biggl[ - s(w) {h'}^2 + r(w)
\partial_{\alpha} h\partial^{\alpha} h \biggr],
\label{pac1}
\end{equation}
where $h$ represents, as usual, a generic tensor polarization.

By using the Euler-Lagrange equations, Eq. (\ref{graveq}) can be obtained.
In terms of $\mu$
the action of Eq. (\ref{pac1}) can be written as 
\begin{equation}
\delta^{(2)} S= \frac{1}{2} \int d^{d + 2} x \biggl[ - {\mu'}^2 - 
\frac{{(\sqrt{s}})''}{\sqrt{s}} \mu^2 + \frac{r}{s} 
\partial_{\alpha}\mu\partial^{\alpha}\mu
\biggr].
\end{equation}
By taking the functional derivative of the above action with 
respect to  $\mu$ the related equations of motion is, as expected, 
Eq. (\ref{canvar}).

The obtained result has been derived without specifying the 
background geometry but only assuming the form of the metric, i.e. 
Eq. (\ref{metric}). Equations (\ref{graveq})--(\ref{canvar}) 
hold then in general for the theory described 
by the action (\ref{ac}). From Eq. (\ref{canvar}), the equation for the mass 
eigenstates is 
\begin{equation}
- \frac{d^2\mu_{{\rm m}}}{dw^2} + V(w) \mu_{{\rm m}} = m^2 \frac{r}{s} 
\mu_{\rm m}, 
\label{eigten}
\end{equation}
where 
\begin{equation}
V(w) = {\cal L}^2 - {\cal L}',\,\,\,\,\, {\cal L} = - 
\frac{(\sqrt{s})'}{\sqrt{s}}.
\end{equation}
In a context of supersymmetric quantum mechanics ${\cal L}$ is the 
superpotential \cite{sus}. In term of the superpotential 
Eq. (\ref{eigten}) can be written in terms of two first order differential 
operators. From this form of the equation for the mass eigenstates, it follows 
that if the lowest mass eigenstate is normalizable the spectrum does not 
contain tachyonic modes.

The lowest mass eigenstate related to Eq. (\ref{graveq}) is given by 
\begin{equation}
\mu_0(w) \simeq \sqrt{s(w)} =  a^{d/2} \sqrt{ 1 - 
2 \epsilon \frac{{\cal H}^2}{a^2}(d-2)(d-1) },
\end{equation}
and the corresponding normalization integral can now 
be written, assuming the background is invariant under
 $w\rightarrow - w$ symmetry
\begin{equation}
2\int_{0}^{\infty} a^d(w) q(w) ~ dw = 
2 \int_{0}^{\infty} a^d(w) \biggl[  1 - 
2  \epsilon \frac{{\cal H}^2}{a^2}(d-2)(d-1)\biggr] ~ dw,
\label{norm}
\end{equation}
where the first equality follows from the definition of Eq. (\ref{q}).
Concerning eq. (\ref{norm}) few remarks are in order.
Consider 
for sake of concreteness the case $d=3$. It will now be shown that 
{\em if} the four-dimensional Planck mass is finite and {\em if} 
the background geometry is regular everywhere, {\em then} the 
tensor zero mode will always be localized. 

From Eq. (\ref{rel1}) the four-dimensional Planck  mass 
can be written as 
\begin{equation}
M_{P}^2 = M^3 \int_{0}^{\infty} a^3 \biggl[ q 
+ \frac{8 \epsilon}{a^2}( {\cal H}'+ {\cal H}^2) \biggr]~ dw. 
\label{mp1}
\end{equation}
Using now the fact that $a({\cal H}^2 + {\cal H}' ) = a''$, Eq. (\ref{mp1})
can be further modified 
\begin{equation}  
M_{P}^2 = M^3  \biggl\{\int_{0}^{\infty} a^3~ q~ dw\biggr\} + 8 \epsilon M^3 
[a']_{0}^{ \infty}
\label{mp2}
\end{equation}
The second term in Eq. (\ref{mp2}) should be finite if the geometry is 
regular everywhere. The first term is exactly proportional 
to the integral appearing in the normalization condition 
of the tensor zero mode of Eq. (\ref{norm}). Hence, if the four-dimensional
Planck mass is finite, the tensor zero mode is localized.

Notice that, if $q(w)$ would diverge 
at some value of $w$, then a singularity 
should be expected in the background. Suppose, in fact, that 
$q(w)$ diverges at some $w$.
If this is the case, from Eq. (\ref{vph}) 
it is  easy to show that 
in order to have $\varphi'$ regular, $({\cal H}' -{\cal H}^2)$ 
should go to zero. But ${\cal H}' \sim {\cal H}^2$ implies that ${\cal H}\sim 
(w_0 - w)^{-1}$, which produces, in its turn, a singularity 
in the curvature invariants (\ref{r1})--(\ref{r3}). Thus, if $q(w)$ at some 
value of $w$ either $\varphi'(w)$ diverges or the curvature invariant 
diverge, or both. 

\renewcommand{\theequation}{5.\arabic{equation}}
\setcounter{equation}{0}
\section{Localization of vector and scalar modes}

\subsection{The vector modes}
The evolution of the gauge-invariant vector fluctuation $V_{\mu}$ can 
can be obtained, after some algebra, from the $(\mu,\nu)$ and $(\mu,w)$ 
component of Eq. (\ref{fl1}). The detailed results for the fluctuations 
of the Riemann and Lanczos tensor perturbed in the amplitude 
of the gauge-invariant vector modes of the geometry are reported in 
Appendix C. From the $(\mu,\nu)$ and $(\mu,w)$ components of Eq. (\ref{fl1})
the resulting equations are, respectively, 
\begin{eqnarray}
&& q(w) [\partial^{(\nu}
V_{\mu)}]' + \biggl\{ d {\cal H} - \frac{ 2\epsilon}{a^2} (d-1) (d-2) 
[ 2 {\cal H} {\cal H}' + (d-2) {\cal H}^3]\biggr\} V_{\mu} =0,
\label{veq1}\\
&& -\partial_{\alpha} \partial^{\alpha} V_{\mu} =0.
\label{veq2}
\end{eqnarray}
Eq. (\ref{veq1}) is obtained from Eqs. (\ref{pveq1})--(\ref{pveq3}) 
whereas Eq. (\ref{veq2}) can be derived 
 from Eqs. (\ref{pveq1a})--(\ref{pveq3a}). Both sets of equations 
are collected in Appendix C.
Eq. (\ref{veq2}) implies that $V_{\mu}$ is massless whereas 
Eq. (\ref{veq1}) allows to determine the evolution of the zero mode.
Recalling the definition of the background function $q(w)$ given in Eq. 
(\ref{q}), Eq. (\ref{veq2}) can be written as 
\begin{equation}
{\cal V}_{\mu}' + \biggl( \frac{d}{2} {\cal H} + \frac{q'}{ 2 q} \biggr) 
{\cal V}_{\mu}=0, 
\end{equation}
where ${\cal V}_{\mu} = a^{d/2} \sqrt{q} V_{\mu}$ is the 
canonical normal mode of the vector action perturbed to second order 
in the amplitude of the amplitude of the metric fluctuations, namely
\begin{equation}
\delta^{(2)} S_{V} = \int d^{d + 1} x d w 
\frac{1}{2}\biggl[ \eta^{\alpha\beta} 
\partial_{\alpha} {\cal V}^{\mu} \partial_{\beta} {\cal V}_{\mu} \biggr].
\label{canvec}
\end{equation}
For each vector polarization, the 
evolution of the zero mode is given by:
\begin{equation}
{\cal V}_0 \simeq \frac{1}{a^{d/2} \sqrt{q}}.
\end{equation}
The corresponding normalization 
integral will then be 
\begin{equation}
2 \int_{0}^{\infty} \frac{d w}{a^d(w) q(w)}. 
\label{norm2}
\end{equation}
By comparing Eq. (\ref{norm2}) with Eq. (\ref{norm}) it can be immediately 
appreciated that the integrands appearing in the two expressions are one the 
inverse of  the other.  
{\em If}  the tensor modes are localized, {\em then} the vector modes
 will not be localized. 

\subsection{ The scalar modes}
Using the definition of $q(w)$ the evolution equations for the gauge-invariant 
scalar fluctuations of the geometry can be written from the explicit 
expressions of Eqs. (\ref{fl1})--(\ref{fl6}). The details 
are reported in Appendix C. For the $(w,w)$ component the results is
\footnote{For sake of simplicity, in the following part and in the 
related Appendix D natural gravitational units $2 \kappa = 1$ will be 
used.}   
\begin{eqnarray}
&& q \partial_{\alpha} \partial^{\alpha} \Xi + ( d + 1) q \Psi'' 
+ ( d + 1) {\cal H} \biggl[ 1 - \frac{ 2 \epsilon}{a^2} ( d-1) ( d-2) ( 
2 {\cal H}' - {\cal H}^2) \biggr] + ( d + 1) {\cal H} q \Xi'
\nonumber\\
&&  + 
2 ( d+ 1)\Xi \biggl\{ {\cal H}' + \frac{\varphi'^2}{2} - 
\frac{2 \epsilon}{a^2} 
( d- 1) ( d - 2) \biggl[ 2 {\cal H}' {\cal H}^2 - {\cal H}^4\biggr]\biggr\}
+ \varphi' X' + \frac{1}{d} \frac{\partial V}{\partial \varphi} a^2 X=0,
\label{ww}
\end{eqnarray}
whereas the $(\mu, \nu)$ component leads to  
\begin{eqnarray}
&& q \Psi'' + {\cal H} \Psi' \biggl\{ ( 2 d + 1)  - \frac{ 2 \epsilon}{a^2} 
(d-1) (d-2) [ 2 {\cal H}' + ( 2 d -1 ) {\cal H}^2)]\biggr\} + 
q {\cal H} \Xi' - q \partial_{\alpha} \partial^{\alpha} \Psi 
\nonumber\\
&&+ 2\Xi 
\biggl\{ ({\cal H}' + d {\cal H}^2 )- \frac{2 \epsilon}{a^2} (d-1) (d-2) 
[( d-1) {\cal H}^4 + 2 {\cal H}' {\cal H}^2 ]\biggr\}  + \frac{1}{d} 
\frac{\partial V}{\partial\varphi} a^2 X=0,
\label{munu}
\end{eqnarray}
if $\mu = \nu$, and to 
\begin{equation}
q \Xi - ( d- 1) \Psi \biggl\{ 1 - \frac{2 \epsilon}{a^2} ( d-2) [
2 {\cal H}'  + (d-3) {\cal H}^2 ]\biggr\} =0.
\label{numu}
\end{equation}
if $\mu \neq \nu$. Finally, the $(\mu, w)$ component of Eq. (\ref{fl1}) 
produces the constraint
\begin{equation}
 d q (\Psi' + {\cal H} \Xi) + \frac{1}{2} \varphi' X =0.
\label{muw}
\end{equation}
The perturbed scalar field equation becomes
\begin{equation}
X'' + d{\cal H} X' - \partial_{\alpha}\partial^{\alpha} X 
- \frac{\partial^2 V}{\partial\varphi^2}a^2 X
+ \varphi'[(d + 1) \Psi' + \Xi'] + 2 ( \varphi'' + d {\cal H} \varphi' ) 
\Xi =0.
\end{equation}

Subtracting now Eq. (\ref{munu}) from Eq. (\ref{ww}) 
\begin{equation}
q \partial_{\alpha} \partial^{\alpha} \biggl( \Psi + \Xi\biggr) + 
d ~ q \biggl[ \Psi'' + {\cal H} \biggl(\Xi' - \Psi'\biggr)\biggr] 
+ d~ q'( \Psi' + {\cal H} \Xi) + X' \varphi' =0.
\label{com}
\end{equation}
Using now Eq. (\ref{numu}) in order to eliminate $\Xi$ from the constraint, 
Eq. (\ref{muw}) can be written as 
\begin{equation}
X = - \frac{2 q d}{\varphi'} \biggl\{ \Psi' + \Psi \biggl[ (d -1) {\cal H} + 
\frac{q'}{q}\biggr] \biggr\}.
\label{con2}
\end{equation}
Inserting Eqs. (\ref{numu})--(\ref{con2}) into Eq. (\ref{com}) 
a decoupled equation for the Bardeen potential is obtained 
\begin{eqnarray}
&&\Psi'' - \biggl[ 1 + \frac{q'}{{\cal H} q} \biggr] \partial_{\alpha}
 \partial^{\alpha} \Psi 
+ \Psi' \biggl[ d {\cal H} + 2 \frac{q'}{q} - 
2 \frac{\varphi''}{\varphi'} \biggr]
\nonumber\\
&&+ \Psi \biggl[ \frac{q''}{q} + \frac{q'}{q} \frac{{\cal H}'}{{\cal H}}
+( d -1)  {\cal H} \frac{ q'}{q} - 2( d- 1) {\cal H} 
\frac{\varphi''}{\varphi'} 
- 2 \frac{\varphi''}{\varphi'} \frac{q'}{q} + 2 ( d - 1) {\cal H}' \biggr]=0.
\label{ps}
\end{eqnarray}
Rescaling now $\Psi$ according to
\begin{equation}
\Phi = \frac{ a^{d/2} q}{\varphi'}\Psi, 
\end{equation}
the following equation can be obtained from Eq. (\ref{con2}) 
\begin{equation}
\Phi'' - z \biggl(\frac{1}{z}\biggr)'' \Phi 
- \biggl( 1 + \frac{ q'}{{\cal H} q} \biggr) \partial_{\alpha}  
\partial^{\alpha} 
\Phi =0,
\label{ph}
\end{equation}
where
\begin{equation}
z(w) = \frac{a^{d/2} \varphi'}{{\cal H}}.
\label{z}
\end{equation} 
In order to get to eq. (\ref{ph}) the following background relation 
[obtained by deriving Eq. (\ref{vph})] has been used
\begin{equation}
\frac{{\cal H}''}{\cal H} = 2 {\cal H}' - 
\biggl(2 \frac{\varphi''}{\varphi'} - \frac{q'}{q} \biggr)
\biggl[ {\cal H} - \frac{{\cal H}'}{\cal H} \biggr]. 
\end{equation}
By virtue of the constraint (\ref{muw}) the equation obeyed by $\Phi$ is also 
obeyed by the appropriately rescaled $\Xi$ variable. 

The normalization condition for the lowest mass eigenstate 
related to Eq. (\ref{ph}) can be now written as 
\begin{equation}
2\int_{0}^{\infty} \frac{dw}{|z(w)|^2}\equiv \int_{0}^{\infty}d w 
\frac{ {\cal H}^2}{a^d ~\varphi'^2} . 
\label{norm3}
\end{equation}
It will now be shown that this integral cannot be convergent. 
Consider the case $d=3$.
The integrand appearing in Eq. (\ref{norm3}) can be 
rearranged by using Eq. (\ref{vph}) which should always hold 
since it is a background relation. Therefore, 
\begin{equation}
\int_{0}^{\infty} d w~\frac{{\cal H}^2}{ a^{d} {\varphi'}^2} 
= 6 \int_{0}^{\infty}\frac{d w}{a^d q}~
\biggl(\frac{{\cal H}^2}{{{\cal H}^2 -{\cal H}'}}\biggr).
\end{equation}
The expression appearing in the denominator is $a^{- 3} q^{-1}$. This is 
exactly the inverse of the integrand arising in the definition of the Planck 
mass. Now, in Eqs. (\ref{mp1}) and (\ref{mp2}) it has been shown that 
in order to have a finite four-dimensional Planck mass $a^3 q$ 
should be convergent everywhere and, in particular, at infinity. Therefore 
$a^{-3} q^{-1}$ will be strongly divergent.
In the same limit, if the curvature invariant are regular at infinity,
${\cal H}^2/( {\cal H}^2 - {\cal H}')$ cannot converge fast enough to make 
the whole integral convergent.  

\renewcommand{\theequation}{6.\arabic{equation}}
\setcounter{equation}{0}
\section{Physical examples}
In the previous sections various conditions on the localization of the 
zero modes of the geometry have been obtained without 
specifying the background solution. It is interesting 
to apply them to some specific case of (analytical) thick brane solution 
with EGB corrcetions.

In order to solve Eqs. (\ref{ex1})--(\ref{ex3}) a useful approach 
is to fix the geometry. Then, from Eq. (\ref{vph}), the relation
$\varphi(w) $ can be obtained after one integration. Then, by 
inverting this relation, $w=w(\phi)$ can be inserted in Eq. (\ref{V}) and 
$V(\varphi)$ determined. Following this approach it can be obtained, 
for instance,
\begin{eqnarray}
&& a(x) = \frac{a_0}{\sqrt{ x^2 + 1}},  
\label{a}\\
&& \varphi(x) = \varphi_0 \frac{ x}{\sqrt{x^2 + 1}} + \varphi_1, 
\label{varph}\\
&& V(\varphi) = V_0 \biggl[ \biggl( \frac{ d + 3}{2} \biggr) 
\biggl( \frac{\varphi -\varphi_1}{\varphi_{0}}\biggr)^4 - ( d + 3)
\biggl( \frac{\varphi-\varphi_1}{\varphi_{0}}\biggr)^2 + 1 \biggr],
\label{potV}
\end{eqnarray}
where 
\begin{eqnarray}
&& a_0 = \sqrt{ 2\, \epsilon\,( d-1)\,( d-2)}\, b,
\label{a0}\\
&& \varphi_0 = \sqrt{\frac{d}{\kappa}} ,\,\,\,
V_0= \frac{d}{4\,\kappa\, \epsilon\,(d-1)\,(d-2)},
\label{ph0}
\end{eqnarray}
and $\varphi_1$ is an integration constant.
In Eqs. (\ref{a})--(\ref{ph}),  $x= b\, w$ is the bulk coordinate rescaled
through the brane thickness. In this dimension-less coordinate 
the brane core is located  for $|x| \leq 1$. Far from the core, 
i.e. $|x|\gg 1$, the limit of the solution is $AdS_{d+2}$ space, in fact 
\begin{equation}
\lim_{|x| \rightarrow \infty } a(x) = \frac{A}{w} 
\end{equation}
where $A$  is the ``radius'' of $AdS$ space. For instance, in the 
$d=3$ case, $A= 2\sqrt{2\alpha'\kappa}$ can be identified with the 
radius of the $AdS_{5}$ space. Similar solutions have been discussed 
in \cite{kak}.

Over these solutions 
the curvature invariants and the EGB combination (appearing in the quadratic 
part of the action) are all regular for any value of the bulk
coordinate. In fact, Eqs. (\ref{r1})--(\ref{r3}) in the specific 
background provided by Eqs. (\ref{a})--(\ref{a0}), the explicit form of 
the curvature invariants can be obtained: 
\begin{eqnarray}
 R^2 &=& \frac{{\left( d + 1\right) }^2\,
    {\left[ \left(  d+ 2 \right) \,x^2 -2 \right] }^2}
    {4\,{\left(  d - 2 \right) }^2\,
    {\left( d -1\right) }^2\,{\epsilon}^2\,
    {\left( x^2 +1 \right) }^2},
\\
 R^{A B} R_{A B} &=&
\frac{\left( d + 1\right) \,
    \left[ d^2\,x^4 + 2\,{\left( x^2 -1\right) }^2 + 
      d\,\left( 3\,x^2-1\right)\left(x^2 -1\right)  \right] }
    {4\,{\left( d -2 \right) }^2\,
    {\left(  d -1\right) }^2\,{\epsilon}^2\,
    {\left( x^2 +1 \right) }^2},
\\
R^{ A B C D} R_{A B C D} &=& 
\frac{\left( d+ 1 \right) \,
    \left[ 2 - 4\,x^2 + \left( 2 + d \right) \,x^4 \right] 
    }{2\,{\left(  d-2 \right) }^2\,
    {\left(  d -1\right) }^2\,{\epsilon}^2\,
    {\left( x^2 + 1\right) }^2}.
\end{eqnarray}
All the curvature invariants go to constant for large $|x|$ and they have 
two minima and a maximum around the origin.

Let us now focus our attention to the case $d=3$. 
In this case 
the integral appearing in the definition of the Planck mass 
[see Eq. (\ref{rel1})]
is finite and the integrand always convergent, as it can be directly 
checked using Eq. (\ref{a}) into Eq. (\ref{rel1}). 

As argued in the Section IV the tensor zero mode 
will also be normalizable. In fact, for the background of Eqs. 
(\ref{a})--(\ref{V}) 
the normalization condition of eq. (\ref{norm}) reads 
\begin{equation}
2 \int_{0}^{\infty} \frac{dw}{ a^3(w) q(w)} \equiv 16 b^2 \epsilon^{3/2}
\int_{0}^{\infty} \frac{d x}{ ( 1 + x^2)^{5/2}}.
\end{equation}
which is clearly convergent.
Consider now the vector modes discussed in Section V. 
On the basis of the results of Eq. (\ref{norm2}) 
the vector zero mode is not normalizable. In fact, from Eq. (\ref{norm2}) 
the normalization integral can be reduced to 
\begin{equation}
\int_{0}^{\infty} ( 1 + x^2)^{5/2} dx
\end{equation}
which is not convergent at infinity. 

Consider finally the scalars discussed in Section V.
From the 
general condition derived in Eq. (\ref{norm3}), the 
normalization integral reduces to 
\begin{equation}
2\int_0^{\infty} \frac{d w}{|z(w)|^2} = 
\frac{ 1}{4 b^3 \epsilon^{3/2} \varphi_0^2}
 \int_{0}^{\infty} x^2 ( 1 + x^2)^{5/2},
\end{equation}
which is divergent at infinity.

\section{Concluding remarks} 

A gauge-invariant framework for the analysis of the fluctuations of 
the various modes of a warped geometry has been developed 
in the case when the gravity action 
contains quadratic curvature corrections parametrized in the Gauss-Bonnet form.

General conditions for the localization of the tensor, vector and scalar 
fluctuations of the metric have been derived. If physical  domain-wall
solutions are described using a bulk action containing the Gauss-Bonnet
combination 
the evolution equations of the fluctuations can be reduced to a set of 
decoupled second order (Schr\"odinger-like) differential equations.

The lowest mass eigenstate of the tensor fluctuations is always 
localized provided the four-dimensional Planck mass is finite. 
On the contrary, 
under the same assumption, the scalar and vector zero modes are not localized. 
The only other assumptions  used in the analysis are that the geometry is 
fully regular and that the background is symmetric for $w\rightarrow - w$.
Specific examples of thick  domain-wall solutions 
induced by Gauss-Bonnet self-interactions have been also studied providing, in 
this way, concrete examples of the general features illustrated in the
analysis of the fluctuations. 

\section*{Acnowledgements}
It is a pleasure to thank M. E. Shaposhnikov for important discussions.

\newpage
\begin{appendix}

\renewcommand{\theequation}{A.\arabic{equation}}
\setcounter{equation}{0}
\section{Explicit form of quadratic invariants} 
Useful background relations will be now collected. From the explicit 
form of the Riemann tensors 
\begin{eqnarray}
&& \overline{R}^{w}_{~~\mu w\nu} = {\cal H}' \eta_{\mu\nu}, 
\nonumber\\
&& \overline{R}^{\mu}_{~~\alpha\nu\beta} = [\delta^{\mu}_{\nu} 
\eta_{\alpha\beta} - 
\delta^{\mu}_{\beta} \eta_{\alpha\nu}] {\cal H}^2,
\end{eqnarray}
the corresponding explicit form  of the Ricci tensors
\begin{eqnarray}
&&\overline{R}_{\mu\nu} = ( {\cal H}' + d {\cal H}^2)\eta_{\mu\nu},
\nonumber\\
&& \overline{R}_{ww} = - ( d+ 1) {\cal H}',
\end{eqnarray}
leads to the explicit expression of the curvature invariants
\begin{eqnarray}
 \overline{R}^2 &=&\frac{(d+1)^2}{a^4}\biggl[ 4 {{\cal H}'}^2 + d^2 {\cal H}^4 
+ 4 d {\cal H}' {\cal H}^2 \biggr] ,
\label{r1}\\
 \overline{R}^{A B} \overline{R}_{A B} 
&=& \frac{( d + 1)}{a^4} \biggl[ (d + 2) {{\cal H}'}^2 
+ d^2 {\cal H}^4 + 2 d {\cal H}' {\cal H}^2 \biggr], 
\label{r2}\\
\overline{R}^{ A B C D} 
\overline{R}_{A B C D} &=& \frac{(d + 1)}{a^4} \biggl[ 4 {{\cal H}'}^2 
+ 2 d {\cal H}^4 \biggr],
\label{r3}
\end{eqnarray}
whereas the explicit form of the EGB combination is 
\begin{equation}
{\cal R}_{\rm EGB}^2 = \frac{d(d+ 1)}{a^4}\biggl[ (d-1) (d-2) {\cal H}^4 + 
4 (d-1) {\cal H}' {\cal H}^2\biggr].
\label{r4}
\end{equation}
The covariant components of the Lanczos tensor are instead:
\begin{eqnarray}
&& \overline{{\cal Q}}_{\mu\nu} = \frac{d (d-1) (d-2) }{2~a^2}
\{(d-3) {\cal H}^4 + 2 {\cal H}^2 {\cal H}'\}\eta_{\mu\nu},
\label{l1}\\
&& \overline{{\cal Q}}_{ww} = - \frac{d (d+ 1) (d-1) (d -2)}{2 a^2} {\cal H}^4.
\label{l2}
\end{eqnarray}

\renewcommand{\theequation}{B.\arabic{equation}}
\setcounter{equation}{0}
\section{The tensor problem}
In the present Appendix the perturbed form of the various quantities 
entering Eqs. (\ref{fl1})--(\ref{fl6}) will be reported in the case 
of the tensor modes of the geometry. The only non-vanishing fluctuations 
of the Riemann and Ricci tensors can be written, in this case, 
as 
\begin{eqnarray}
&& \di R^{\nu}_{~~ w \mu w} = - 
( {h_{\mu}^{\nu}}'' + {\cal H} {h_{\mu}^{\nu}}'),
\label{ptr1}\\
&& \di R^{\alpha}_{~~\mu\beta\nu} = \partial_{\beta} [ - 
\partial^{\alpha} h_{\mu\nu} + \partial_{\mu} h^{\alpha}_{\nu} + 
\partial_{\nu} h_{\mu}^{\alpha}] - \partial_{\nu}[ - 
\partial^{\alpha} h_{\mu\beta} \partial_{\mu} h_{\beta}^{\alpha} + 
\partial_{\beta} h_{\mu}^{\alpha}] 
\nonumber\\
&&+ {\cal H}[ \eta_{\mu\nu} 
{h^{\alpha}_{\beta}}' - \eta_{\mu\beta} {h_{\nu}^{\alpha}}' ]
+ {\cal H}[ \delta^{\alpha}_{\beta} ( h_{\mu\nu}' + 2 {\cal H} h_{\mu\nu})
- \delta_{\nu}^{\alpha} ( h_{\mu\beta}' + 2 {\cal H} h_{\mu\beta})],
\label{ptr2}\\
&& \di R^{\nu}_{\mu} = \frac{1}{a^2} ( {h_{\mu}^{\nu}}'' + d {\cal H} 
{h_{\mu}^{\nu}}' - \partial_{\alpha}\partial^{\alpha} h_{\mu}^{\nu} ).
\label{ptr3}
\end{eqnarray}
Hence the different contributions appearing in Eq. (\ref{peq}) can 
be written in an explicit form and they are
\begin{eqnarray}
&& [ 1 - 2 \epsilon \overline{R}] \di R_{\mu}^{\nu} = 
\frac{1}{a^2} \biggl\{ 1 - \frac{ 2 \epsilon}{a^2} \biggl[ 2 {\cal H}' ( d-1) 
+ d( d-3) {\cal H}^2\biggr]\biggr\} 
( {h_{\mu}^{\nu}}'' + d {\cal H} {h_{\mu}^{\nu}}'
- \partial_{\alpha} \partial^{\alpha} {h_{\mu}^{\nu}}), 
\label{peq1}\\
&& 4\epsilon [ \di R_{\mu\alpha} \overline{R}^{\alpha\nu} + 
\overline{R}_{\mu\alpha} \di R^{\alpha\nu}] = \frac{4\epsilon}{a^4} 
\biggl\{ {h_{\mu}^{\nu}}'' [ (d + 1) {\cal H}' - {\cal H}^2] + 
 {h_{\mu}^{\nu}}'[ 2 d {\cal H} {\cal H}' + d ( d - 2) {\cal H}^3]
\nonumber\\
&&- [{\cal H}' + (d -1) {\cal H}^2 ] \partial_{\alpha}\partial^{\alpha} 
{h_{\mu}^{\nu}}
\biggr\},
\label{peq2}\\
&& 2 \epsilon [ 2 \di R_{\alpha\beta} 
\overline{ R}_{\mu}^{~~\alpha\nu\beta} + 2 \overline{R}_{\alpha\beta} 
\di R_{\mu}^{~~\alpha\nu\beta} 
+ 2 \overline{R}_{ww} \di R_{\mu}^{~~w\nu w} 
- \di R_{\mu C D E} \overline{R}^{\nu C D E} 
\nonumber\\
&& - 
\overline{R}_{\mu C D E}\di R^{\nu C D E}]=  - \frac{ 8 \epsilon}{a^4}
\biggl\{ {\cal H}' {h_{\mu}^{\nu}}'' + [{\cal H} {\cal H}' + ( d- 1) 
{\cal H}^3 ] {h_{\mu}^{\nu}}' - {\cal H}^2 \partial_{\alpha} \partial^{\alpha} 
{h_{\mu}^{\nu}}\biggr\}
\label{peq3}
\end{eqnarray}
Inserting the terms reported iun Eqs. (\ref{peq1})--(\ref{peq3}) 
into Eq. (\ref{peq}) the explicit form of the evolution 
of the tensor modes of the geometry can  be obtained and it is given in Eq. 
(\ref{graveq}).

\renewcommand{\theequation}{C.\arabic{equation}}
\setcounter{equation}{0}
\section{The vector problem}
The gauge-invariant fluctuations of the Riemann tensors to first order in the 
amplitude of the vector perturbations  of the metric will now be reported
\begin{eqnarray}
&& 
\di R^{\mu}_{~~w \nu w} 
= \frac{1}{2}\biggl\{\partial_{\nu}[ {V^{\mu}}' +{\cal H}
V^{\mu}]+ \partial^{\mu}[V_{\nu}'+ {\cal H} V_{\nu}]\biggr\},
\label{v1}\\
&&
\di R^{\mu}_{~~\alpha w \nu} = ({\cal H}^2 -{\cal H}') 
V^{\mu} \eta_{\alpha\beta} 
- {\cal H}^2 V_{\alpha} \delta^{\mu}_{\beta} + 
\frac{1}{2} \partial_{\beta}[\partial^{\mu} V_{\alpha} 
- \partial_{\alpha}V^{\mu}],
\label{v2}\\
&& \di R^{\alpha}_{~~\mu\beta\nu} = \frac{{\cal H}}{2} 
\biggl\{ 2 \partial_{\nu} 
V^{\alpha}\eta_{\mu\beta} - \partial_{\beta} V^{\alpha} \eta_{\mu\nu} - 
\partial^{\alpha} V_{\beta} \eta_{\mu\nu} - \delta_{\beta}^{\alpha} ( 
\partial_{\mu} V_{\nu} + \partial_{\nu} V_{\mu})
\nonumber\\
&& - 
\eta_{\mu\beta} [\partial_{\nu} V^{\alpha} - \partial^{\alpha}V_{\nu}]
+ \delta_{\nu}^{\alpha} [\partial_{\mu} V_{\beta} + \partial_{\beta} V_{\mu} ]
\biggr\}.
\label{v3}
\end{eqnarray}
Using Eqs. (\ref{v1})--(\ref{v3}),  the explicit form of Eq. (\ref{fl1}) 
perturbed to first order in the amplitude of the vector fluctuations of the 
metric can be obtained. In particular, 
for the $(\mu,\nu)$ component\footnote{Notice that, in the following 
equations, $\partial_{(\mu} V_{\nu)}
=\frac{1}{2} ( \partial_{\mu} V_{\nu} + \partial_{\nu} V_{\mu} )$}
\begin{eqnarray}
&& [ 1 - 2 \epsilon \overline{R}] \di R_{\mu}^{\nu} = 
\nonumber\\
&&-\frac{1}{a^2} \biggl[ 1 - \frac{2\epsilon}{a^2}\biggl(2 (d+ 1) {\cal H}' 
+ d(d+1) {\cal H}^2 \biggr) \biggr]
\biggl\{ \bigl[ \partial_{(\mu}V^{\nu)}\bigr]' + d {\cal H} 
\bigl[\partial_{(\mu}V^{\nu)}\bigr] \biggr\},
\label{pveq1}\\
&& 4\epsilon [ \di R_{\mu\alpha}~ \overline{R}^{\alpha\nu} + 
\overline{R}_{\mu\alpha} \di R^{\alpha\nu}] = 
- \frac{8\epsilon}{a^4}({\cal H}' + d{\cal H}^2)\biggl\{ 
\biggl[\partial_{(\mu} V^{\nu)}\biggr]' + d {\cal H} \partial_{(\mu} V^{\nu)}
\biggr\},
\label{pveq2}\\
&& 2 \epsilon [ 2 \di R_{\alpha\beta} 
\overline{ R}_{\mu}^{~~\alpha\nu\beta} + 2 \overline{R}_{\alpha\beta} 
\di R_{\mu}^{~~\alpha\nu\beta} + 2 \overline{R}_{ww} \di 
R_{\mu}^{~~w\nu w} 
\nonumber\\
&& - \di R_{\mu C D E} \overline{R}^{\nu C D E} -
\overline{R}_{\mu C D E}\di R^{\nu C D E}]= 
\nonumber\\
&& \frac{4 \epsilon}{a^4} \biggl\{ [ {\cal H}^2 - ( d-1){\cal H}'] 
\biggl[ \partial^{(\nu}V_{\mu)} \biggr]' -  
[ (d (d -2) -2 (d-1)) {\cal H}^3  + 2 ( d -1) {\cal H} {\cal H}']\biggl[
 \partial^{(\nu}V_{\mu)}
\biggr]\biggr\}.
\label{pveq3}
\end{eqnarray}
For the $(\mu, w)$ component we have 
\begin{eqnarray}
&& [ 1 - 2 \epsilon\overline{R}] ~\di R_{\mu}^{w} = 
\frac{2}{ a^2} \partial_{\alpha} \partial^{\alpha} V_{\mu} \biggl\{1 
- \frac{2 \epsilon}{a^2} [ 2 ( d+ 1) {\cal H}' + d ( d + 1) {\cal H}^2 ]
{\cal H}^2\biggr\},
\label{pveq1a}\\
&& 4\epsilon [ \di R_{\mu w}~ \overline{R}^{w w} + 
\overline{R}_{\mu\alpha} \di R^{\alpha w}] = \frac{ 4 \epsilon}{2 a^4} 
\partial_{\alpha} \partial^{\alpha} V_{\mu} [ d {\cal H}^2 + 
( d + 2) {\cal H}'], 
\label{pveq2a}\\
&& 2 \epsilon [ 2 \di R_{C D}~ 
\overline{ R}_{\mu}^{~~C w D} + 2 \overline{R}_{C D}~
\di R_{\mu}^{~~C w D}  
- \di R_{\mu C D E}~ \overline{R}^{\nu C D E} 
\nonumber\\
&&-
\overline{R}_{\mu C D E}~\di R^{\nu C D E}]= 
\frac{ 2 \epsilon}{a^4}[ (d -1){\cal H}^2 - {\cal H}' ] \partial_{\alpha}
\partial^{\alpha} V_{\mu}. 
\label{pveq3a}
\end{eqnarray}
Using Eqs. (\ref{pveq1})--(\ref{pveq3}), Eq. (\ref{veq1}) can be derived. 
Similarly, Eqs. (\ref{pveq1a})--(\ref{pveq3a}) give Eq. (\ref{veq2}).
 
\renewcommand{\theequation}{D.\arabic{equation}}
\setcounter{equation}{0}
\section{The scalar problem}

The gauge-invariant fluctuations of the Riemann tensors to first order 
in the scalar fluctuations of the metric 
\begin{eqnarray}
&&\di R^{w}_{~~\mu w \nu} = \eta_{\mu\nu} [ \Psi'' + {\cal H} 
( \Psi' + \Xi') + 2 {\cal H}' ( \Psi + \Xi)],
\label{scp1}\\
&& \di R^{\mu}_{~~\alpha w \beta} = [\delta_{\mu}^{\beta}\partial_{\alpha} 
\Psi' - \eta_{\alpha\beta} \partial^{\mu} \Psi'] - {\cal H}[ \partial^{\mu} \Xi
\eta_{\alpha\beta} - \delta_{\beta}^{\mu} \partial_{\alpha} \Xi],
\label{scp2}\\
&& \di R^{\alpha}_{~~\mu\beta\nu} = \delta_{\nu}^{\alpha} \partial_{\mu}
\partial_{\beta} \Psi - 
\delta_{\beta}^{\alpha} \partial_{\mu}\partial_{\nu} - \eta_{\mu\nu} 
\partial_{\beta}\partial^{\alpha} \Psi + \eta_{\mu\beta} 
\partial_{\nu}\partial^{\alpha} \Psi 
\nonumber\\
&&+ [ 2 {\cal H} \Psi' + 2 {\cal H}^2 ( \Psi + \Xi)][ \eta_{\mu\nu} 
\delta^{\alpha}_{\beta} 
- \eta_{\mu\beta} \delta_{\nu}^{\alpha} ],
\end{eqnarray}
allow to compute the various components of the perturbed form of Eqs. 
(\ref{fl1}).
The $(\mu, w)$ component of Eq. (\ref{fl1}) can be obtained 
from the following expression 
\begin{eqnarray}
&& \di R_{\mu}^{w} - \frac{1}{2} \partial_{\mu}\chi\partial^{w}\varphi = 
\frac{1}{a^2} \biggl[ \partial_{\mu} \biggl( d \Psi' + d {\cal H} \Xi - 
\frac{1}{2} \varphi' \chi\biggl)
\biggr],
\label{sceq1}\\
&& \epsilon \biggl[ - 2  \overline{R} \di R_{\mu}^{w} + 
4 \di R_{\mu w} \overline{R}^{w w} + 4
\overline{R}_{\mu\alpha} \di R^{\alpha w} \biggr] =
\frac{2\epsilon}{ a^2} [ 2 {\cal H}' - 2 d^2 {\cal H}^2] \partial_{\mu}[ 
d \Psi' + d {\cal H} \Xi ] ,
\label{sceq2}\\
&& 4 \epsilon \biggl[ \di R_{w \alpha} \overline{R}_{\mu}^{~~w w \alpha} +
\overline{R}_{\alpha\beta} \di R_{\mu}^{~~\alpha w \beta} \biggr]= \frac{
 4 \epsilon}{a^2} d {\cal H}^2 
\partial_{\mu} [ \Psi' + d {\cal H} \Xi],
\label{sceq3}\\
&& - 2 \epsilon \biggl[ \overline{R}_{\mu\alpha\beta\gamma} 
\di R^{w \alpha\beta\gamma} + 
\di R_{\mu\alpha\beta\gamma} \overline{R}^{w \alpha w \beta} + 
\di R_{\mu\alpha\beta w} \overline{R}^{ w \alpha \beta w} \biggr]=
\nonumber\\
&& \frac{4 \epsilon}{a^4} ( {\cal H}' - {\cal H}^2) \partial_{\mu}[ d \Psi' 
+ d {\cal H} \Xi].
\label{sceq4}
\end{eqnarray}
The EGB combination appearing in Eq. (\ref{fl1}) does not contribute 
to the off-diagonal components of the perturbed equations
but it does contribute to the diagonal components. 
From the fluctuation of the EGB combination
\begin{eqnarray}
&& {\cal K}= 
\epsilon \di {\cal R}^2_{\rm EGB} \equiv \frac{4 d\epsilon}{a^4} 
\biggl\{ (d -1) {\cal H}^2 \partial_{\alpha}
\partial^{\alpha}\Xi -  (d-1) [ {\cal H}' + (d -2) {\cal H}^2] 
\partial_{\alpha}\partial^{\alpha} \Psi 
\nonumber\\
&& +  (d-1) ( d+ 1) {\cal H}^2 \Psi'' + (d -1) (d + 1) {\cal H}^3 \Xi' 
+ (d-1) ( d + 1) {\cal H} \Psi' [ 2 {\cal H}' + (d -1) {\cal H}^2] 
\nonumber\\
&& +
( d-1) (d + 1) [ 4 {\cal H}' {\cal H}^2 + (d -2) {\cal H}^4] \Xi \biggr\},
\label{pgb}
\end{eqnarray}
the  $(w,w)$ and $(\mu,\nu)$ components of Eq. (\ref{fl1}) can be obtained. 
 In fact, defining
\begin{eqnarray}
 {\cal M}_{A}^{B} &=& \di R_{A}^{B} - \frac{1}{2} \partial_{A}
 \varphi \partial^{B} X + \frac{1}{d} 
\frac{\partial V}{\partial\varphi} X \delta_{A}^{B},
\nonumber\\
 {\cal N}_{ A}^{B} &=& - 2 \epsilon \biggl[ \overline{R} \di R_{A}^{B} +
 \overline{R}_{A}^{B} \di R \biggr],
\nonumber\\
 {\cal O}_{A}^{B} &=&  4 \epsilon\biggl\{ \di R_{A C} \overline{R}^{B C} 
+ \overline{R}_{A C } \di R^{A C} \biggr\},
\nonumber\\
{\cal S}_{A}^{B} &=&  4\epsilon \biggl\{ \di R_{C D} 
\overline{R}_{A}^{~~C B D} +
 \overline{R}_{C D} 
\di R_{A}^{~~C B D } \biggr\},
\nonumber\\
{\cal T}_{A}^{B} &=& - 4 \epsilon \biggl\{ \overline{R}_{A C D E} 
\di R^{ B C D E} + 
\di R_{A C D E} \overline{R}^{B C D E} \biggr\},
\label{defp}
\end{eqnarray}
the equation of the fluctuations is given, according to Eq. (\ref{fl1}), by :
\begin{equation}
{\cal M}_{A}^{B} + \frac{1}{d} \delta_{A}^{B} {\cal K} + {\cal N}_{A}^{B} + 
{\cal O}_{AB} + 
{\cal S}_{A}^{B} + {\cal T}_{A}^{B} =0
\label{sum}
\end{equation}
Hence, the $(w, w)$ equation can be derived from the following explicit 
expressions
\begin{eqnarray}
{\cal M}_{w}^{w} &=& \frac{1}{a^2} 
\biggl\{ \partial_{\alpha}\partial^{\alpha} \Xi + ( d + 1) [ \Psi'' + 
{\cal H} ( \Psi' + \Xi')] + [2 ( d + 1) {\cal H}' \varphi'^2] \Xi + 
\varphi' X' 
\nonumber\\
&& + 
\frac{1}{d} \frac{\partial V}{\partial\varphi} a^2 X\biggr\},
\label{sceq7}\\ 
{\cal N}_{w}^{w} &=& - \frac{2\epsilon}{a^4} \biggl\{ ( d+ 1) ( 4 {\cal H}'
 + d {\cal H}^2) 
 \partial_{\alpha} \partial^{\alpha} \Xi  
- 2 d (d+ 1) {\cal H}' \partial_{\alpha}\partial^{\alpha} \Psi
\nonumber\\
&& + 4( d + 1)^2[ 2 {\cal H}'^2+ {\cal H}^2 {\cal H}'] \Xi  
 +( d + 1)^2[ 4 {\cal H}' + d {\cal H}^2] (\Psi''+ {\cal H} \Xi') 
\nonumber\\
&&+(d + 1)^2 [ 2 ( d + 2) {\cal H}' + d {\cal H}^2]{\cal H} \Psi' \biggr\},
\label{sceq6}\\
{\cal O}_{w}^{w} &=& \frac{4\epsilon}{a^4} ( d + 1) {\cal H}' \biggl\{ 
2 \partial_{\alpha} \partial^{\alpha} \Xi + 
2 ( d + 1) [ \Psi'' + {\cal H} ( \Psi' + \Xi')] + 4 ( d + 1) {\cal H}' 
\Xi \biggr\},
\label{sceq8}\\
{\cal S}_{w}^{w} &=& \frac{4 \epsilon}{a^4} \biggl\{ ( d + 1) [ d {\cal H}^2 
+ 2 {\cal H}' ] \Psi'' 
+ ( d + 1) [ 2 ( d + 1) {\cal H}' + d {\cal H}^2] {\cal H} \Psi' 
- 2 d {\cal H}' \partial_{\alpha} \partial^{\alpha} \Psi
\nonumber\\
&& +4 ( d + 1) {\cal H}'   ( {\cal H}' + d {\cal H}^2)  \Xi
+ ( d + 1) [ 2 {\cal H}' + d {\cal H}' ] {\cal H} \Xi'
( 2 {\cal H}' + d {\cal H}^2) \partial_{\alpha} \partial^{\alpha} \Xi \biggr\},
\label{sceq9}\\
{\cal T}_{w}^{w} &=& - \frac{ 8 \epsilon}{a^4} {\cal H}' \biggl\{ 
(d + 1) [ \Psi '' + {\cal H} ( \Psi' + \Xi' ) + 2 {\cal H}' \Xi ] 
+ \partial_{\alpha}\partial^{\alpha} \Xi + 
\partial^{\alpha} \partial^{\alpha} \Xi \biggr\}.
\label{sceq10}
\end{eqnarray}
Using Eqs. (\ref{sceq7})--(\ref{sceq10}) together with (\ref{pgb}) into 
Eq. (\ref{sum}) 
Eq. (\ref{ww}) is recovered. 

The $(\mu,\nu)$ component can be similarly obtained from
\begin{eqnarray}
{\cal M}_{\mu}^{\nu} &=& \frac{1}{a^2} \biggl\{ \delta_{\mu}^{\nu} \biggl[ 
\Psi'' + ( 2 d + 1) {\cal H} \Psi' + 
{\cal H} \Xi' + 2 ( {\cal H}' + d {\cal H}^2) \Xi - \partial_{\alpha} 
\partial^{\alpha} \Psi 
\nonumber\\
&&+ \frac{1}{d} \frac{\partial V}{\partial\varphi} a^2 X \biggr] + 
\partial_{\mu}\partial^{\nu}[ \Xi - 
( d - 1) \Psi],
\label{sceq11}\\
{\cal N}_{\mu}^{\nu} &=& - \frac{2 \epsilon}{ a^4} \biggl\{ \delta_{\mu}^{\nu}
 \biggl[
  ( d + 1) [ 2 ( 3 d + 2 ) {\cal H}' + d ( 4 d + 3)  
{\cal H}^2 ]{\cal H} \Psi' 
\nonumber\\
&&  +  (d+ 1) ( 4 {\cal H}' + 3 d {\cal H}^2)\Psi +
4 ( d + 1) [ 2 {\cal H}'^2 + d^2 {\cal H}^4 +  3 d {\cal H}' {\cal H}^2] \Xi  
\nonumber\\
&& + ( d + 1) [ 4 {\cal H}' + 3 d {\cal H}^2] {\cal H}\Xi'  
- [ 2 ( 2 d + 1) {\cal H}' 
+ d ( 3 d + 1) {\cal H}^2] \partial_{\alpha} \partial^{\alpha} \Psi 
\nonumber\\
&& + 2 ( {\cal H}' + d {\cal H}^2) \partial_{\alpha}\partial^{\alpha} \Xi 
\biggr]  
+( d+ 1) [ 2 {\cal H}' + d {\cal H}^2] \partial_{\mu}\partial^{\nu} [ \Xi 
- ( d-1 ) \Psi] \biggr\},
\label{sceq12}\\
{\cal O}_{\mu}^{\nu} &=& \frac{8 \epsilon}{a^4} \biggl\{ \delta_{\mu}^{\nu} ( 
{\cal H}' + d {\cal H}^2) \biggl[ 
\Psi'' + ( 2 d + 1) {\cal H} \Psi' + {\cal H} \Xi'
\nonumber\\
&&  + 2 ( {\cal H}' + d {\cal H}^2) \Xi - 
\partial_{\alpha}\partial^{\alpha} \Psi \biggr] + 
({\cal H}'  + d {\cal H}^2) \partial_{\mu}\partial^{\nu}[ \Xi - ( d -1) \Psi] 
\biggr\}
\label{sceq13}\\
{\cal S}_{\mu}^{\nu} &=& \frac{4 \epsilon}{a^4} \biggl\{ \delta_{\mu}^{\nu} 
\biggl[ ( 2 ( d + 1) {\cal H}' + d {\cal H}^2) \Psi'' 
+ ( 2 ( 2 d + 1) {\cal H}' + d ( 4 d + 1) {\cal H}^2) {\cal H} \Psi'
\nonumber\\
&& + ( {\cal H}' + {\cal H}^2) \partial_{\alpha} \partial^{\alpha} \Xi 
+ [ 4 ( d + 1) {{\cal H}'}^2 + 4 d {\cal H}' {\cal H}^2 + 4 d^2 {\cal H}^4 ] 
\Xi
\nonumber\\
&& - [ ( 3d - 1 ) {\cal H}^2 + {\cal H}' ] \partial_{\alpha} 
\partial^{\alpha} \Psi 
+ [ d {\cal H}^2 + 2 (d + 1) {\cal H}'] {\cal H} \Xi'
\biggr] 
\nonumber\\
&&+
\partial_{\mu} \partial^{\nu} \biggl[ \Xi \biggl( ( d + 1) {\cal H}' 
- {\cal H}^2 \biggr) - 
( d - 1) \biggl( ( d -1 ) {\cal H}^2 + {\cal H}'\biggr) \Psi \biggr],
\label{sceq14}\\
{\cal T}_{\mu}^{\nu} &=& 
 - \frac{ 2 \epsilon}{a^4} \biggl\{ \delta_{\mu}^{\nu} \biggl[ 4 {\cal H}' 
\Psi'' 
+ 4 {\cal H} 
{\cal H}' ( \Psi' + \Xi') + 8 ( {\cal H}'^2 + d {\cal H}^4 ) \Xi
\nonumber\\
&& - 4 {\cal H}^2 \partial_{\alpha} \partial^{\alpha} \Psi+
8 d {\cal H}^3 \Psi'\biggr]
+ \partial_{\mu}\partial^{\nu} \biggl[ 4 {\cal H}' \Xi - 4 ( d - 1) {\cal H}^2 
\Psi\biggr]\biggr\}.
\label{sceq15}
\end{eqnarray}
Again using Eqs. (\ref{sceq11})--(\ref{sceq15}) together with Eq. (\ref{pgb}) 
into Eq. (\ref{sum}), Eqs. (\ref{munu}) and (\ref{numu}) 
(reported in the text)  
are recovered.

\end{appendix}

\end{document}